\documentclass[aps,twocolumn,prb,preprintnumbers,amsmath,amssymb,superscriptaddress,floats]{revtex4}

\usepackage{amsmath}	
\usepackage{gensymb}
\usepackage{hyperref}
\hypersetup{colorlinks=true}
\usepackage{upgreek}
\usepackage{graphics}
\usepackage{hyperref}
\usepackage{epsfig}
\usepackage{color}
\usepackage{bm}
\usepackage{float}
\usepackage{ulem} 

\usepackage{CJK}        
\usepackage{url}
\usepackage{multirow}

\newcommand{\upperRomannumeral}[1]{\uppercase\expandafter{\romannumeral#1}}

\begin{document}

\title{Intrinsic Jump Character of the First-Order Quantum Phase Transitions}
\author{Qiang Luo}
\email[]{qiangluo@ruc.edu.cn}
\affiliation{Department of Physics, Renmin University of China, Beijing 100872, China}
\author{Jize Zhao}
\email[]{zhaojz@lzu.edu.cn}
\affiliation{School of Physical Science and Technology $\&$ Key Laboratory for Magnetism and
Magnetic Materials of the MoE, Lanzhou University, Lanzhou 730000, China}
\author{Xiaoqun Wang}
\email[]{xiaoqunwang@sjtu.edu.cn}
\affiliation{Key Laboratory of Artificial Structures and Quantum Control (Ministry of Education), School of Physics and Astronomy, Tsung-Dao Lee Institute, Shanghai Jiao Tong University, Shanghai 200240, China}
\affiliation{Collaborative Innovation Center for Advanced Microstructures, Nanjing 210093, China}
\affiliation{Beijing Computational Science Research Center, Beijing 100084, China}

\date{\today}

\begin{abstract}
  We find that the first-order quantum phase transitions~(QPTs) are characterized by intrinsic jumps of relevant operators
  while the continuous ones are not.
  Based on such an observation, we propose a bond reversal method
	where a quantity $\mathcal{D}$, the difference of bond strength~(DBS), is introduced
  to judge whether a QPT is of first order or not.
  This method is firstly applied to
  an exactly solvable spin-$1/2$ \textit{XXZ} Heisenberg chain and a quantum Ising chain with longitudinal field
  where distinct jumps of $\mathcal{D}$ appear at the first-order transition points for both cases.
  We then use it to study the topological QPT of a cross-coupled~($J_{\times}$) spin ladder where the Haldane--rung-singlet transition switches from being continuous to exhibiting a first-order character at $J_{\times, I} \simeq$ 0.30(2).
  Finally, we study a recently proposed one-dimensional analogy of deconfined quantum critical point
  connecting two ordered phases in a spin-$1/2$ chain.
  We rule out the possibility of weakly first-order QPT because the DBS is smooth when crossing the transition point.
  Moreover, we affirm that such transition belongs to the Gaussian universality class with the central charge $c$ = 1.
\end{abstract}


\maketitle
\textit{Introduction}.--
Understanding how strongly correlated systems order into different phases
as well as the transitions among them
remains one of the most fundamental and significant problems
in modern condensed matter physics\cite{SachdevBook_2011,LMMBook_2011,WenBook_2019}.
In particular, the quantum phase transitions~(QPTs), which occurr at zero temperature,
are omnipresent phenomena and could in general be classified into two types.
One is continuous when the ground state of the system changes continuously at the transition point,
accompanied by a diverging correlation length and vanishing energy gap.
The other is, instead, of first order when the order parameter and other relevant observables
display discontinuity across the transition point.
While traditional continuous QPTs are well described by the Landau-Ginzburg-Wilson~(LGW) theory,
recent years have witnessed some exceptions such as
topological QPT\cite{KT_1973,TsuiSG_1982,Haldane_1983} and
deconfined quantum critical point~(DQCP)\cite{SenthilVBetal2004,SenthilBSetal2004,Sandvik2007,ShaoGS2016}
that beyond the scope of LGW paradigm.
The topological QPT may take place between two disordered phases,
thus it can not be detected by local order parameters\cite{PollmannTBetal_2010,FaureBCVO_2018}.
What's more, it could be
either continuous or of first order upon a fine-tuned interaction strength
\cite{Wang_2000,AmaricciBCetal_2015,BarbarinoSB_2019,RoyGS_2016}.
The DQCP was proposed by Senthil \textit{et. al.}\cite{SenthilVBetal2004,SenthilBSetal2004}
whereby a continuous QPT occurs between two spontaneously symmetry breaking~(SSB) phases
and the critical point implies an emergent symmetry.
The $J$-$Q$ model\cite{Sandvik2007} is such an example
where extensive numerical studies provide evidences for a continuous (or weakly first order) transition
between a N{\'e}el phase and a valence bond solid~(VBS) phase\cite{Sandvik2010,ChenHDetal2013,IaizziDS2018}.

In contrast to the continuous QPTs, the first-order QPTs are less studied so far
despite they appear frequently in quantum many-body systems.
Remarkably, a first-order QPT called a photon-blockade breakdown was observed experimentally
in a driven circuit quantum electrodynamics system\cite{FinkDVetal2017}.
The finite-size scaling of gap and various probes borrowed from quantum information sciences near the transition points of the first-order QPTs have been discussed until recently\cite{LaumannMSetal_2012,MuellerJJ_2014,CampostriniNPetal_2014,YusteCCetal_2018,RossiniVicari_2018}.
Therefore, it is of vital importance to devise appropriate tools for a proper characterization of their dominating features.

Let us consider a Hamiltonian of the form\cite{SachdevBook_2011}
\begin{equation}\label{HamH0H1}
\mathcal{H}(\lambda) = \mathcal{H}_0 + \lambda\mathcal{H}_I
\end{equation}
where $\lambda$ is a driving parameter.
If $\mathcal{H}_0$ and $\mathcal{H}_I$ commute,
then both of them could be simultaneously diagonalized and eigenfunctions are independent of $\lambda$. %
This means that the spectra of $\mathcal{H}_0$ and $\mathcal{H}_I$ are irrelevant of $\lambda$,
while the total ground-state energy $E_g(\lambda) = \langle \mathcal{H}(\lambda)\rangle$ could vary linearly with $\lambda$.
Consequently, there can be a level-crossing point $\lambda_t$
where the ground-state energy per site $e_g(\lambda)$ = $(1/L)E_g(\lambda)$ = $e_0 + \lambda e_I$ as a function of $\lambda$ exhibits nonanalyticity.
Here, $L$ is the number of lattice sites.
It should be aware that continuous QPT could also occur and we move the discussion to the supplemental material~(SM)\cite{SuppMat}.
Taking the derivative of $e_g(\lambda)$ with respect to $\lambda$,
a jump of $e_I$ at $\lambda_t$ will appear, indicating of a first-order QPT.
The jump also reflects the structural change of ground-state wave function.
Because of the continuity of energy $e_g(\lambda)$, a similar jump for $e_0$ is also expected to eliminate the singularity.
To detect the transition point $\lambda_t$,
a quantity $\mathcal{D}$ dubbed the difference of bond strength~(DBS) is introduced to magnify the jump behaviors.
It is defined as
\begin{equation}\label{DBSDef}
\mathcal{D} = e_0 - \textrm{sgn}(\lambda_t)e_I,
\end{equation}
where the minus sign reflects the spirit of the bond reversal method.
Nevertheless, in most systems the two terms $\mathcal{H}_0$ and $\mathcal{H}_I$ do not commute,
resulting in a cumbersome expression of $e_g(\lambda)$ versus $\lambda$.
However, the main spirit remains unchanged in that
the jump character of $\mathcal{D}$ faithfully inherits the discontinuity of first-order QPTs.

In what follows we will firstly illustrate the bond reversal method
in two different but thoroughly studied one-dimensional~(1D) spin models
that possess first-order QPTs:
(i) a celebrated spin-$1/2$ Heisenberg \textit{XXZ} chain for which
all the energy as well as the DBS $\mathcal{D}$ can be calculated analytically and
(ii) a quantum Ising chain with both longitudinal and transverse fields which does not host exact solution but the transition line is well-known.
Having established the cornerstone of our method,
we then apply it to
(iii) a topological QPT of a cross-coupled spin ladder and (iv) a recently proposed spin-$1/2$ chain with DQCP,
both of which are beyond the scope of the conventional LGW paradigm.
All the models are studied by the density-matrix renormalization group~(DMRG) method
\cite{White_1992,White_1993,PeschelBook_1999,Schollwoeck_2005},
which is a powerful tool for dealing with quantum-mechanical problems in 1D systems.
We utilize the periodic boundary condition~(PBC) for the first two cases to have a better comparison with analytical results.
For the latter cases, however, we turn to the open boundary condition~(OBC) which is beneficial to large-scale numerical calculations.

\textit{XXZ chain}.--
The 1D spin-$1/2$ Heisenberg \textit{XXZ} chain has long served as the workhorse for the study of quantum magnetism\cite{SchollwockBook_2004}.
Its Hamiltonian is given by
\begin{align}\label{Ham-XXZ}
\mathcal{H} = \sum_{i=1}^{L} \frac12\left(S_i^{+}S_{i+1}^{-} + S_i^{-}S_{i+1}^{+}\right) + \Delta S_i^{z}S_{i+1}^{z}
\end{align}
where $S_i^{\pm} = S_i^{x} + iS_i^{y}$ is the raising/lowering operator at site $i$
and $\Delta$ is the anisotropic parameter.
In particular, in the region $-1 < \Delta \leq 1$
the ground state is a Luttinger liquid~(LL) with a gapless excitation spectrum.
The ground-state energy per site $e_g$ in the thermodynamic limit~(TDL) $L\to\infty$
can be calculated as\cite{YangYang1966,ShiroishiTakahashi2015,CloizeauxGaudin1966}
\begin{equation}\label{BetheAnsatzGSenergy}
e_g^{\textrm{LL}}(\Delta) = \frac{\Delta}{4} - \frac{\sin\pi\upsilon}{\pi}\int_0^{\infty}\Big(1-\frac{\tanh\upsilon x}{\tanh x}\Big)\textrm{d}x
\end{equation}
with $\Delta = \cos\pi\upsilon$.
Beyond the critical region it presents a long-range ferromagnetic (FM) or antiferromagnetic (AFM) order,
exhibiting in correspondence to the FM point $\Delta = -1$ a first-order QPT,
and a continuous one belonging to the Kosterlitz-Thouless~(KT) universality class
at the AFM point $\Delta = 1$.
In the FM phase ($\Delta < -1$), the spins are parallel along the $z$ direction,
resulting in $e_g^{\textrm{FM}} = \Delta/4$.
The spin-spin correlation functions could be obtained by Hellmann-Feynman theorem.
In the LL phase, however, no simple expressions for the correlation functions are available
except for some rational $\upsilon$-values\cite{BortzSS2007}.
Historically, the explicit expressions of the correlation functions\cite{SuppMat}
were first given by Jimbo and Miwa in 1996\cite{JimboMiwa1996},
and then simplified by Kato~\textit{et al.} several years later\cite{KatoSTetal2003}.
According to Eq.~\eqref{DBSDef}, the DBS is defined as
$\mathcal{D}_L$ = $\langle S_{L/2}^zS_{L/2+1}^z\rangle$ + $2\langle S_{L/2}^xS_{L/2+1}^x\rangle$.

\begin{figure}[!ht]
\centering
\includegraphics[width=0.95\columnwidth, clip]{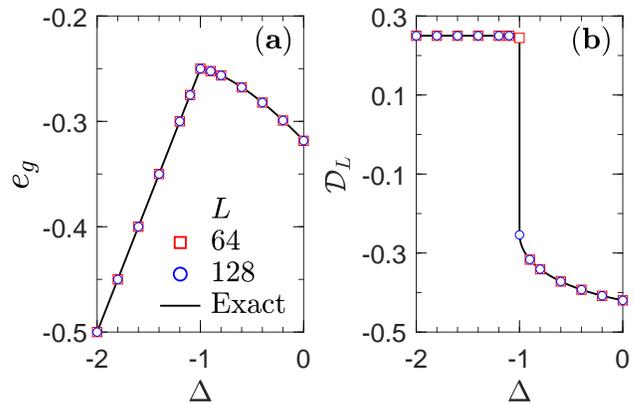}\\
\caption{(a) The ground-state energy $e_g$ for $L$ = 64~(red rhombus), 128~(blue circle), and TDL (black line).
(b) The same setup as (a) for DBS $\mathcal{D}_L$.}\label{FIG-XXZ}
\end{figure}

At $\Delta = -1$ the Hamiltonian Eq.~\eqref{Ham-XXZ} posses a hidden FM $SU(2)$ symmetry,
resulting in a ground-state manifold of degenerate $SU(2)$ multiplet corresponding to the largest total spin.
The model is not conformal invariant and dramatic changes of
its entanglement behaviors occur\cite{BanchiCV2009,ErcolessiEFetal2011,AlbaHL2012,StasinskaRPetal2014}.
In Fig.~\ref{FIG-XXZ}(a) we show the ground-state energy $e_{g}$ around the transition point $\Delta = -1$.
A vivid cusp of the energy curve could be spotted at $\Delta = -1$,
while it turns to be a jump of $\mathcal{D}_L$ in Fig.~\ref{FIG-XXZ}(b).
This gives clearly a first glance of the jump character of $\mathcal{D}$ in the first-order QPT\cite{SuppMat}.

\begin{figure}[!ht]
\centering
\includegraphics[width=0.95\columnwidth, clip]{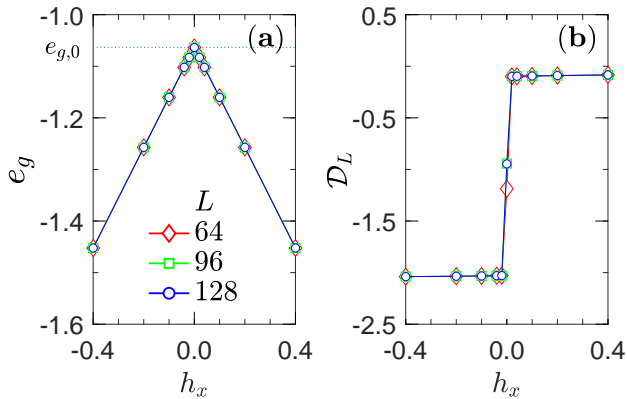}\\
\caption{(a) The ground-state energy $e_g$ for $L$ = 64~(red rhombus), $L$ = 96~(geren square), and 128~(blue circle).
(b) The same setup as (a) for DBS $\mathcal{D}_L$.}\label{FIG-TFIM}
\end{figure}

\textit{Quantum Ising chain with longitudinal field}.--
The 1D quantum Ising chain is integrable and its exact solution was presented by Pfeuty in 1970\cite{Pfeuty_1970}.
It owns a continuous QPT of the Ising universality class
separating a FM and a paramagnetic phase at the critical value of transverse field $h_z (\ge 0)$\cite{DuttaACetal_2015}.
What's more, an emergent $E_8$ symmetry was experimentally verified around the critical point\cite{ColdeaTWetal_2010}.
By introducing a longitudinal field $h_x$ the model is no longer integrable
except for a specially fine-tuned weak longitudinal field\cite{Zamolodchikov_1989}.
The total Hamiltonian is thus given by
\begin{align}\label{Ham-IsingHxHz}
\mathcal{H} = -\sum_{i=1}^{L} \left({\sigma}_i^x{\sigma}_{i+1}^x + h_z{\sigma}_i^z + h_x{\sigma}_i^x\right)
\end{align}
where $\hat{\sigma} = (\sigma^x, \sigma^y, \sigma^z)$ are the Pauli matrices.
In the FM phase~($h_z < 1$) of the magnetic phase diagram\cite{YusteCCetal_2018,ODK_2003,AtasBogomolny_2017},
a first-order QPT with a discontinuity of the magnetization, $M_x = \frac1L\sum_i\langle\sigma_i^x\rangle$,
takes place at $h_x = 0$.
Specifically, let $h_z = 1/2$ we have the exact ground-state energy
$e_{g,0}$ = $-\frac{3}{4\pi} E\big(\sqrt{6\sqrt2}/3\big)$ $\approx$ $-1.0635$
where $E(\cdot)$ is the complete elliptic integral of second kind\cite{Pfeuty_1970}.
As can be seen from Fig.~\ref{FIG-TFIM}(a), there is a pinnacle at $h_x = 0$ in the energy curve,
and the symmetric feature is a reminiscence of $\mathbb{Z}_2^x$ symmetry.
The DBS is defined as
$\mathcal{D}_L$ =
$\big\langle{\sigma}_{L/2}^x\big\rangle - \big\langle\big({\sigma}_{L/2}^x{\sigma}_{L/2+1}^x + h_z{\sigma}_{L/2}^z\big)\big\rangle$,
which is merely a shift of $M_x$ by $e_{g,0}$ currently.
In this occasion $\mathcal{D}_L$ plays the role of order parameter $M_x$ and
there is no wonder that it exhibits a jump at $h_x = 0$ (see Fig.~\ref{FIG-TFIM}(b)). 
In general, since DBS has an ambiguous relation with order parameter,
it's thus well founded to regard this jump as a signal for a first-order QPT\cite{SuppMat}.

\textit{Cross-coupled spin ladder}.--
The role of frustration in quasi-1D magnetic materials has attracted numerous attention\cite{SchollwockBook_2004}
ever since the discovery of high-temperature superconductivity in 1980s\cite{BednorzMuller_1986}.
The cross-coupled spin ladder\cite{Xian_1995,Wang_2000,WesselNMH_2017}, in particular, is one of the most outstanding models
which is not only of theoretical importance\cite{WhiteLadder_1996,Vekua_2006,Metavitsiadis_2017}
but also experimentally accessible\cite{Dagotto_1996}.
The Hamiltonian of the model reads as follows:
\begin{align}\label{CCSL}
\mathcal{H} = & J_{\parallel}\sum_{i=1}^{L}\sum_{\alpha=1,2} \mathbf{S}_{i,\alpha}\cdot \mathbf{S}_{i+1,\alpha} + J_{\perp}\sum_{i=1}^{L}\mathbf{S}_{i,1}\cdot \mathbf{S}_{i,2} \nonumber \\
& J_{\times}\sum_{i=1}^{L}\big(\mathbf{S}_{i,1}\cdot \mathbf{S}_{i+1,2} + \mathbf{S}_{i,2}\cdot \mathbf{S}_{i+1,1}\big),
\end{align}
where $\mathbf{S}_{i,\alpha}$ denotes a spin-1/2 operator at site $i$ of the $\alpha$-th leg.
$J_{\parallel} ( = 1)$ and $J_{\perp}$ are the NN interactions along the leg and rung directions, respectively.
$J_{\times} > 0$ is the antiferromagnetic cross-coupled interaction.

Whereas a continuous QPT with a central charge $c = 2$
occurs at $J_{\perp} = 0$ in the absence of $J_{\times}$\cite{HijiiKN_2005},
contentious results with a decade disputing exist for nonzero $J_{\times}$.
On the one hand, a columnar dimerized phase was predicted
between the Haldane phase and the rung-singlet phase in a narrow parameter region
at weak cross-coupled interaction $J_{\times}$\cite{Starykh_2004}.
Though some clues for the dimerized phase appear at finite-size case\cite{Liu_2008,Li_2012},
people now generally believe that there is no such a phase actually
\cite{Hung_2006,Kim_2008,Hikihara_2010,Barcza_2012,ChenCZetal_2016}.
On the other hand,
when $J_{\times} = 1$, the Hamiltonian of Eq.~\eqref{CCSL} undergoes a first-order QPT at $J_{\perp, t} = 1.401484$\cite{Xian_1995}.
Due to the dual symmetry of Eq.~\eqref{CCSL}\cite{WeihongKO_1998},
we shall just concentrate on the case where $J_{\times}$ is below the dual line $J_{\times}$ = 1.
Though various numerical calculations have firmly established that
such a first-order QPT remains present for deviations away from the dual line as large as $J_{\times} = 0.6$,
unanimous conclusion has not been drawn on
whether the first-order QPT could extend to all the locus of the phase boundary
or just end at a nonzero inflection point $J_{\times, I}$.
At weak interchain couplings,
an early analytic result predicted that the transition is always of first order\cite{KimFSetal_2000},
and later a numerical calculation of the same group yields to the conclusion\cite{KimLS_2008}.
Meanwhile, in the work of Wang\cite{Wang_2000},
it is found that the first-order QPT is dismissed at $J_{\times, I} = 0.287$,
and a continuous QPT down to the vanishing interchain couplings takes over afterward.
It's worth mentioning that the fact that a continuous QPT occurs at $J_{\times} = 0.2$ is checked
by tensor network approach\cite{ChenCZetal_2016} and quantum Monte Carlo method\cite{WesselNMH_2017}.
In view of the ambiguity, it is our purpose to determine the inflection point $J_{\times, t}$
accurately by bond reversal method.
\begin{figure}[!ht]
\centering
\vspace{-0.35cm}
\includegraphics[width=0.95\columnwidth, clip]{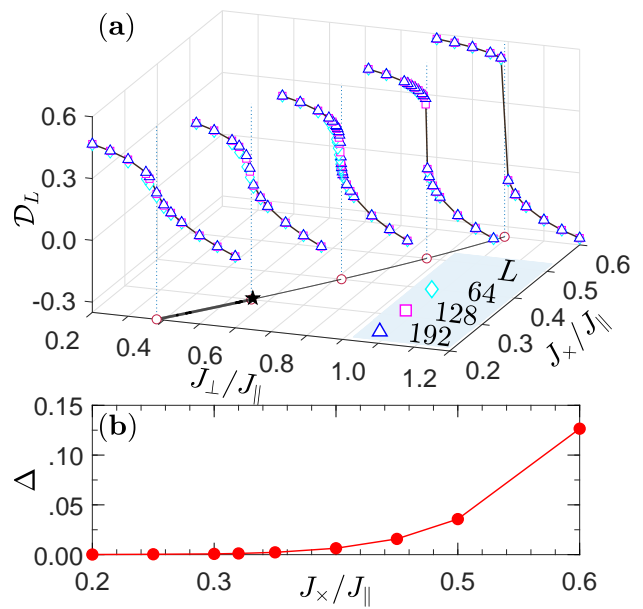}\\
\caption{(a)~The DBS $\mathcal{D}_L$ for $L$ = 64~(cyan rhombus), $L$ = 128~(magenta square), and $L$ = 192~(blue triangular) of different $J_{\times}$'s. The solid lines are guided for the eyes.
In the bottom projection plane, the thick black line and the thin black line are the continuous and first-order phase boundaries, respectively. The pentagram~($\bigstar$) marks the inflection point.
(b)~The Haldane gap $\Delta$ along the phase boundary.}\label{FIG-CCSLDBS}
\end{figure}

During each calculation we shall fix $J_{\times}$ and vary $J_{\perp}$ of Eq.~\eqref{CCSL}.
We thus define the DBS as
$\mathcal{D}_L$ = $\mathcal{R}_L - \mathcal{L}_L$
where $\mathcal{R}_L = \big\langle\big(S^z_{L/2,1}S^z_{L/2,2}+S^z_{L/2+1,1}S^z_{L/2+1,2}\big)\big\rangle$
and $\mathcal{L}_L = \big\langle\big(S^z_{L/2,1} + S^z_{L/2,2}\big)\big(S^z_{L/2+1,1} + S^z_{L/2+1,2}\big)\big\rangle$
in a plaquette in the spirit of Eq.~\eqref{DBSDef}.
In Fig.~\ref{FIG-CCSLDBS}(a) we show the curvatures of DBS for different $J_{\times}$'s from 0.2 to 0.6.
Here, we keep as large as 2000 states typically in our DMRG calculation and extend to 3000 states when necessary.
For $J_{\times}$ = 0.5 and 0.6, there is a jump of DBS in each case, indicating that a first-order QPT occurs.
For other cases that are smaller than $J_{\times}$ = 0.4, however,
the curves are rather smooth and no conspicuous jumps are encountered.
This is a strong evidence that the transitions here are not of first order but continuous with a central charge $c = 2$\cite{SuppMat}.
Whereas the curvatures of DBS for $J_{\times}$ = 0.4 seem to be smooth,
a jump which is a signal for first-order QPT appears for large enough system size\cite{SuppMat}.
We also calculate the energy gap of Haldane phase, i.e.,
$\Delta_L$ = $E_g(S_{\textrm{tot}}^z = 2)$ $-$ $E_g(S_{\textrm{tot}}^z = 0)$,
and the results are shown in Fig.~\ref{FIG-CCSLDBS}(b).
It could be found that the gap is infinitesimal within our numerical precision when $J_{\times} \lesssim 0.30$,
and it opens exponentially afterward.
After a series of careful calculations
we thus conclude that the inflection point $J_{\times, I}$ is a finite value of 0.30(2).

\textit{Spin-1/2 chain with DQCP}.--
Whereas the DQCP was originally proposed in two-dimensional systems\cite{SenthilVBetal2004,SenthilBSetal2004},
the 1D analogy of DQCP was constructed quite recently\cite{JiangMotrunich2019}
and it has been studied by several parallel works on frustrated spin-1/2 chains with discrete symmetries
\cite{RobertsJiangMotrunich2019,HuangLYetal2019,MudryFMetal2019}.
For concrete, we consider the following anisotropic model,
\begin{align}\label{Ham-DQCP}
\mathcal{H} = \sum_{i=1}^{L}\sum_{\upsilon=x,z} -J_{\upsilon}S_i^{\upsilon}S_{i+1}^{\upsilon} + K_{\upsilon}S_i^{\upsilon}S_{i+2}^{\upsilon},
\end{align}
where $J_{\upsilon}, K_{\upsilon} > 0$ so that the NN interactions are ferromagnetic while the 2nd-NN interactions are antiferromagnetic.
We shall treat the NN interaction $J_x = 1$ and fix the 2nd-NN interaction $K = K_{x/z} = 1/2$
so that the only adjustable parameter is $J_z (> 0)$.
When $J_z$ is not very large, the ground state of Eq.~\eqref{Ham-DQCP}
could be continuously connected to that of the Majumdar-Ghosh point\cite{MajumdarGhosh_1969} where $J_z = 1$.
This phase is the well-known dimerized VBS phase which breaks translational symmetry.
On the contrary,
in the regime where $J_z$ is dominant the spins align parallelly along their $z$ directions,
resulting in a $z$FM phase with breaking $\mathbb{Z}_2^z$ symmetry.
In the original work of Jiang \textit{et. al.}\cite{JiangMotrunich2019},
the VBS--$z$FM transition was argued to be continuous,
a transition of which is at odds with the LGW theory
where a direct transition between two states breaking irrelevant symmetries should be of first order.
The critical point $J_{z,c}$ is called DQCP in analogy with its two-dimensional counterpart,
and a continuous $O(2)\times O(2)$ symmetry emerges\cite{JiangMotrunich2019}.
The model Eq.~\eqref{Ham-DQCP} has been studied by matrix product state~(MPS)
which works directly in the TDL in two independent calculations\cite{RobertsJiangMotrunich2019,HuangLYetal2019}.
Both order parameters of the SSB phases have a tiny but finite jump around the critical point.
Notwithstanding, such discontinuity is argued to be an artifact of MPS method.
In fact, the weakly first-order phase transition
is hardly distinguishable from a continuous one\cite{Sandvik2010,ChenHDetal2013,IaizziDS2018},
and thus meticulous calculations should be carried out to check the type of the transition.
We therefore resort to DMRG method where up to 2000 states are kept to revisit this problem.

\begin{figure}[!ht]
\centering
\includegraphics[width=0.95\columnwidth, clip]{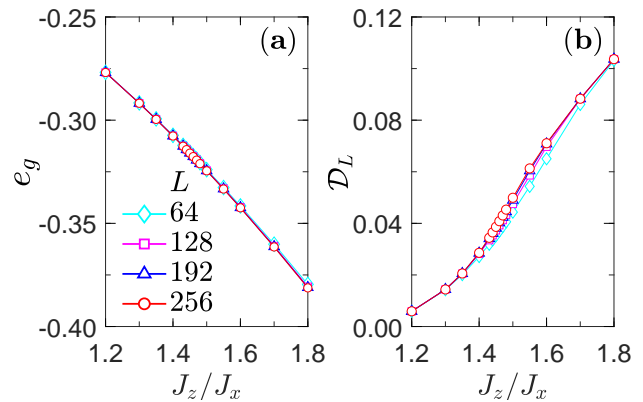}\\
\caption{(a) The ground-state energy $e_g$ for $L$ = 64~(cyan rhombus), $L$ = 128~(magenta square), $L$ = 192~(blue triangular), and 256~(red circle).
(b) The same setup as (a) for DBS $\mathcal{D}_L$.}\label{FIG-DQCPEgDBS}
\end{figure}

To begin with, we calculate the ground-state energy
and the energy curves shown in Fig.~\ref{FIG-DQCPEgDBS}(a) is rather smooth.
The DBS $\mathcal{D}_L = \big\langle S^z_{L/2}S^z_{L/2+1} - S^x_{L/2}S^x_{L/2+1} \big\rangle$
(see Fig.~\ref{FIG-DQCPEgDBS}(b)) is continuous likewise when tuning $J_z$
and no overt jump could be observed in the curves.
This implies that the transition is indeed not a first-order one.

\begin{figure}[!ht]
\centering
\includegraphics[width=0.95\columnwidth, clip]{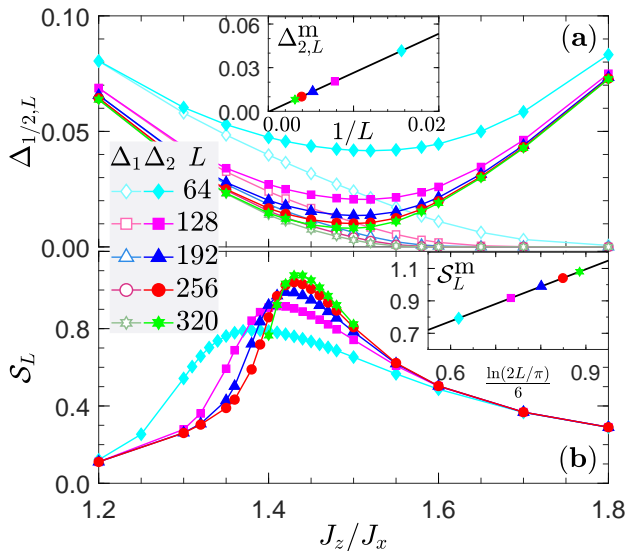}\\
\caption{(a) The first two energy gaps $\Delta_1$~(open symbols) and $\Delta_2$~(filled symbols). The inset shows a linear extrapolation of $\Delta_2$ to TDL. (b) Evolution of vNE $\mathcal{S}_L$. Inset: Logarithmic extrapolation of peaks of $\mathcal{S}_L$ at different length $L$'s.}\label{FIG-DQCPGapCC}
\end{figure}

Because of the OBC utilized in our simulations,
the VBS phase only has a unique ground state while the $z$FM phase still has two-fold degeneracy.
We define the energy gaps $\Delta_{1,2} = E_{1,2}-E_g$,
as the total energy difference between the first/second excited states $E_{1,2}$ and the ground state $E_g$.
In Fig.~\ref{FIG-DQCPGapCC}(a) we show energy gaps $\Delta_{1,2}$ versus $J_z$.
With the increasing of $J_z$, $\Delta_1$ decreases all the way and vanishes rapidly when crossing the critical point.
For $\Delta_2$, However, there is a minimum $\Delta_{2,L}^{\textrm{m}}$ at each length $L$ around the critical point.
As shown in the inset, the $\Delta_{2,L}^{\textrm{m}}$'s follow a linear scaling versus $1/L$
and the gap at TDL is 0.0000(4), indicating the closure of energy gap at the critical point.
Because of the linear scaling ansatz, the critical point is conformal invariant\cite{FrancescoMSBook_1997}.
The von Neumann entropy~(vNE)~$\mathcal{S}_L$ is calculated by the minimal entangled ground state
and the final result is shown in Fig.~\ref{FIG-DQCPGapCC}(b).
A hump appears near the critical point, and this is another evidence for a continuous QPT.
We fit the maxima of vNE $\mathcal{S}_L$ as a function of length $L$,
$\mathcal{S}_L^{\textrm{m}} = \frac{c}{6}\ln\big(\frac{2 L}{\pi}\big) + c'$,
where $c$ is the central charge and $c^\prime$ is a nonuniversal constant\cite{CalabreseCardy2004}.
We find that $c \simeq 1.02(5)$ at the critical point.

We now calculate the critical point $J_{z, c}$ and critical exponents of the order parameters.
The VBS phase is characterized by the difference of the adjacent bond strength, \textit{i.e.},
$M_L^{\textrm{VBS}} = \vert\langle \textbf{S}_i\cdot\textbf{S}_{i+1}\rangle - \langle \textbf{S}_{i-1}\cdot\textbf{S}_{i}\rangle\vert$.
The $z$FM phase has a nonzero local moment at each site and thus $M_L^{z\textrm{FM}} = \vert\langle S_i^z\rangle\vert$.
In practice, we could set $i = L/2$ to minimize the finite-size effect.
Also, when calculating the $M_L^{z\textrm{FM}}$,
a finite pinning field of order 1 is added at the boundaries of the open chain so as to select a determinate ground state.
Theoretically, the order parameter $M_L$ versus $J_z$ with the length $L$ follows\cite{BarberBook_1983}
\begin{equation}\label{FSS}
M_{L}(J_z) \simeq L^{-\beta/\nu}f_M\left(\vert J_z-J_{z,c}\vert L^{1/\nu}\right),
\end{equation}
where the critical exponent $\nu$ describes the divergence of the correlation length
and $\beta$ is the critical exponent of the order parameter
such that $M\sim\vert J_z-J_{z,c}\vert^{\beta}$ near the critical point $J_{z,c}$.
\begin{figure}[!ht]
\centering
\includegraphics[width=0.95\columnwidth, clip]{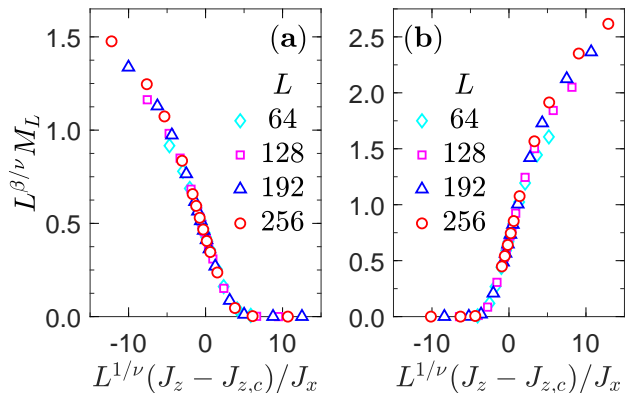}\\
\caption{The FSS of the order parameters (a)~$M_L^{\textrm{VBS}}$ and (b)~$M_L^{z\textrm{FM}}$.
The critical point $J_{z,c}$ and critical exponents $\beta$ and $\nu$ are shown in Tab.~\ref{Tab-DQCPFSS}.}\label{FIG-DQCPFSS}
\end{figure}

In Fig.~\ref{FIG-DQCPFSS} we apply the finite-size scaling~(FSS) method
to the (a)~VBS and (b)~$z$FM phases in the range of $J_z \in [1.20, 1.80]$.
The scaling results are pretty good when $J_z$ is close to the critical point $J_{z,c}$.
Some data, however, deviate from the scaling function when $J_z$ is far away from $J_{z,c}$.
The best fitting values of critical point $J_{z,c}$ and critical exponents are presented in Tab.~\ref{Tab-DQCPFSS}.
The overall critical point $J_{z,c} \simeq 1.4646(6)$,
which is fairly in consistent with previous work by Huang \textit{et. al}\cite{HuangLYetal2019}.
The critical exponents are almost identical for both order parameters, in agreement with the property of the DQCP\cite{JiangMotrunich2019}.
The final results are $\beta = 0.53(3)$ and $\nu = 1.55(6)$.
Interestingly, we find the quantity $2\nu(1-2\beta/\nu)$ equals to 1 roughly,
as predicted from the Luttinger theory where both order parameters
could be expressed by a sole Luttinger parameter\cite{JiangMotrunich2019,RobertsJiangMotrunich2019}.
We also check the cases where $K \neq 1/2$\cite{SuppMat} and indeed find that the critical exponents change with $K$.
Together with the central charge $c \approx 1$,
we could say that the VBS--$z$FM transition belongs to the Gaussian universality class and
the critical point shows some similarities to the LL phase.
\begin{table}[!ht]
\vspace{-0.50cm}
\caption{\label{Tab-DQCPFSS}
Extracted critical point $J_{z,c}$ and corresponding critical exponents $\beta$ and $\nu$ for the continuous VBS--$z$FM phase transition.}
\begin{ruledtabular}
\begin{tabular}{ c | c c c c c c}
Phase  & $J_{z,c}$ & $\beta/\nu$ & $1/\nu$ & $\beta$ & $\nu$ & $2(\nu-2\beta)$  \\
\colrule
VBS    & 1.4647(3)  & 0.344(2) & 0.64(2) & 0.53(2)   & 1.55(5)  & 0.98\\
$z$FM  & 1.4645(5)  & 0.350(3) & 0.65(3) & 0.54(2)   & 1.54(7)  & 0.92
\end{tabular}
\end{ruledtabular}
\end{table}

\textit{Conclusions.}--
In this paper, we propose a bond reversal method to determine a first-order quantum phase transition~(QPT)
by a quantity $\mathcal{D}$ called the difference of bond strength~(DBS).
A first-order QPT could be detected by a jump in DBS and the discontinuity point is exactly the transition point.
The method is rather efficient and could be easily implemented in almost every numerical methods.
We use it to study two unconventional QPTs which are both beyond the scope of Landau-Ginzburg-Wilson theory.
For the cross-coupled ($J_{\times}$) spin ladder,
we clarify that a continuous QPT indeed occurs at weak interchain couplings,
and the inflection point separating the continuous and first-order QPTs is $J_{\times, I} \simeq$ 0.30(2).
For a recently proposed spin-$1/2$ chain which owns two spontaneously symmetry breaking phases,
we confirm that the transition is continuous
because the DBS is fairly smooth and the energy gap vanishes when crossing the critical point.
After a careful finite-size scaling analysis,
we find that the transition belongs to the Gaussian universality class with the central charge $c$ = 1.

\textit{Acknowledgements.}--
We thank S. Hu and S. Jiang for fruitful discussions. 
Q.L. was financially supported by the Outstanding Innovative Talents
Cultivation Funded Programs 2017 of Renmin University of China.
J.Z. was supported by the the National Natural Science Foundation of China (Grant No. 11874188)
and the Fundamental Research Funds for the Central Universities.
X.W. was supported by the National Program on Key Research Project (Grant No. 2016YFA0300501)
and by the National Natural Science Foundation of China (Grant No. 11574200).

\textit{Note added.}--Recently, we became aware of a work on spin-$1/2$ chain with DQCP
that supports our findings\cite{SunWeiKou2019}.

\clearpage

\onecolumngrid

\newpage

\newcounter{equationSM}
\newcounter{figureSM}
\newcounter{tableSM}
\stepcounter{equationSM}
\setcounter{equation}{0}
\setcounter{figure}{0}
\setcounter{table}{0}
\setcounter{page}{1}
\makeatletter
\renewcommand{\theequation}{\textsc{sm}-\arabic{equation}}
\renewcommand{\thefigure}{\textsc{sm}-\arabic{figure}}
\renewcommand{\thetable}{\textsc{sm}-\arabic{table}}


\begin{center}
{\large{\bf Supplemental Material for\\ ``Intrinsic Jump Character of the First-Order Quantum Phase Transitions''}}
\end{center}
\begin{center}
Qiang Luo$^{1}$, Jize Zhao$^{2}$, and Xiaoqun Wang$^{3, 4, 5}$\\
\quad\\
$^1$\textit{Department of Physics, Renmin University of China, Beijing 100872, China}\\
$^2$\textit{School of Physical Science and Technology $\&$ Key Laboratory for Magnetism and\\
Magnetic Materials of the MoE, Lanzhou University, Lanzhou 730000, China}\\
$^3$\textit{Key Laboratory of Artificial Structures and Quantum Control (Ministry of Education),\\ School of Physics and Astronomy, Tsung-Dao Lee Institute,\\ Shanghai Jiao Tong University, Shanghai 200240, China}\\
$^4$\textit{Collaborative Innovation Center for Advanced Microstructures, Nanjing 210093, China}\\
$^5$\textit{Beijing Computational Science Research Center, Beijing 100084, China}
\end{center}

\twocolumngrid



\section{DBS of continuous QPTs}
\subsection{From the second-order Ising transition to the infinite-order KT transition}
\begin{figure}[!ht]
\centering
\includegraphics[width=0.95\columnwidth, clip]{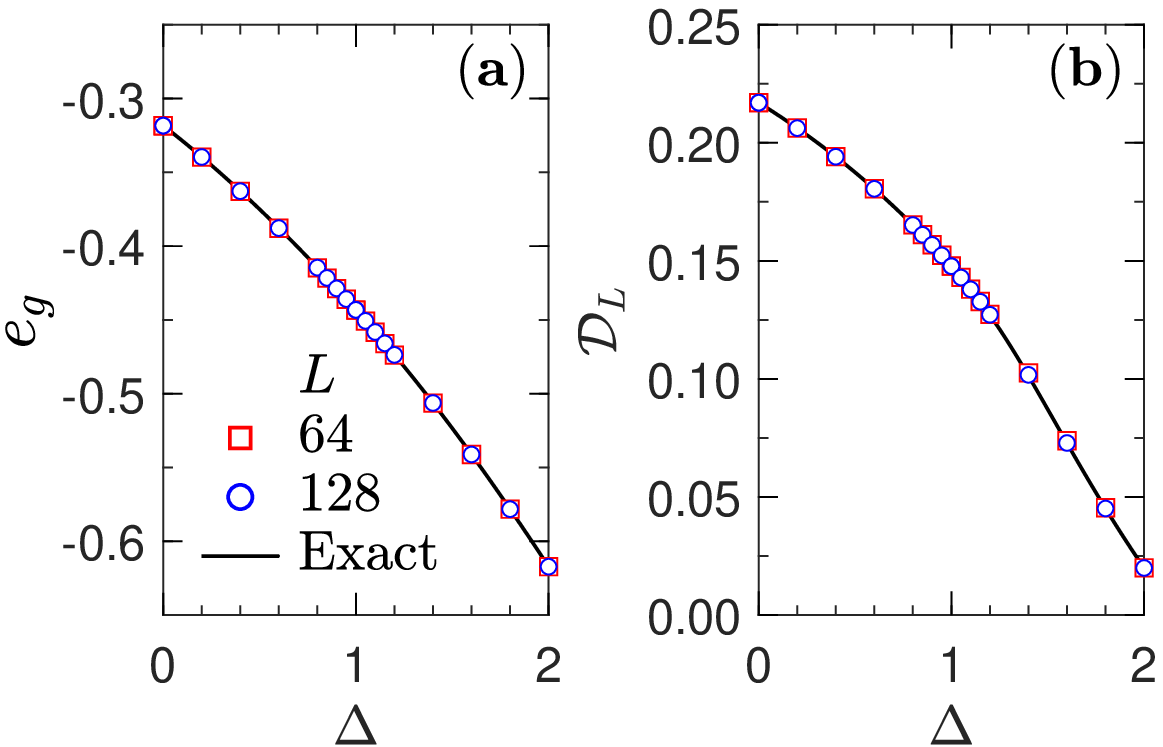}\\
\caption{(a) The ground-state energy $e_g$ of \textit{XXZ} chain for $L$ = 64~(red square), 128~(blue circle), and TDL (black line).
(b) The same setup as (a) for DBS $\mathcal{D}_L$.}\label{FIG-XXZBKT}
\end{figure}

In this section we show the difference of bond strength (DBS) $D$ of continuous QPTs of the KT~(infinite order) and Ising~(second order) universality classes, respectively.
For the spin-1/2 \textit{XXZ} Heisenberg chain,
the transition from the Luttinger liquid phase to the AFM phase
occurs at $\Delta$ = 1 and belongs to the KT universality classes.
In the critical region, i.e, $-1 < \Delta \leq 1$, the correlation functions read as\cite{SMKatoSTetal2003}
\begin{equation}\label{XXZ-LLSx}
\langle S_i^xS_{i+1}^x\rangle = -\frac{1}{4\pi\sin\pi\upsilon}\mathcal{I}_1 + \frac{\cos\pi\upsilon}{4\pi^2}\mathcal{I}_2
\end{equation}
and
\begin{equation}\label{XXZ-LLSz}
\langle S_i^zS_{i+1}^z\rangle = \frac14 + \frac{\cot\pi\upsilon}{2\pi}\mathcal{I}_1 - \frac{1}{2\pi^2}\mathcal{I}_2,
\end{equation}
where $\Delta = \cos\pi\upsilon$ and the integrals
\begin{equation*}
\mathcal{I}_1 \equiv \mathcal{I}_1(\upsilon) = \int_{-\infty}^{\infty} \frac{dx}{\sinh x}\frac{\sinh (1-\upsilon)x}{\cosh \upsilon x}
\end{equation*}
and
\begin{equation*}
\mathcal{I}_2 \equiv \mathcal{I}_2(\upsilon) = \int_{-\infty}^{\infty} \frac{dx}{\sinh x}\frac{x\cosh x}{(\cosh \upsilon x)^2}.
\end{equation*}
In the massive region,  i.e, $\Delta = \cosh\phi > 1$, the correlation functions are\cite{SMTakahashiKS2003}
\begin{equation}\label{XXZ-AFSx}
\langle S_i^xS_{i+1}^x\rangle = \frac{1}{8}\int_{-\infty+i/2}^{\infty+i/2} \frac{dx}{\sinh \pi x}
\frac{x\sinh2\phi - \sin2\phi x}{\sin^2\phi x \sinh\phi}
\end{equation}
and
\begin{equation}\label{XXZ-AFSz}
\langle S_i^zS_{i+1}^z\rangle = \frac14 + \frac{1}{4}\int_{-\infty+i/2}^{\infty+i/2} \frac{dx}{\sinh \pi x}
\frac{\sin2\phi x\coth\phi - 2x}{\sin^2\phi x}.
\end{equation}
The DBS is defined as
$\mathcal{D}_L$ = $\langle S_{L/2}^zS_{L/2+1}^z\rangle - 2\langle S_{L/2}^xS_{L/2+1}^x\rangle$.
and the final result is presented in Fig.~\ref{FIG-XXZBKT}.
It could be found that the DBS is smooth when crossing the critical point $\Delta = 1$.

\begin{figure}[!ht]
\centering
\includegraphics[width=0.95\columnwidth, clip]{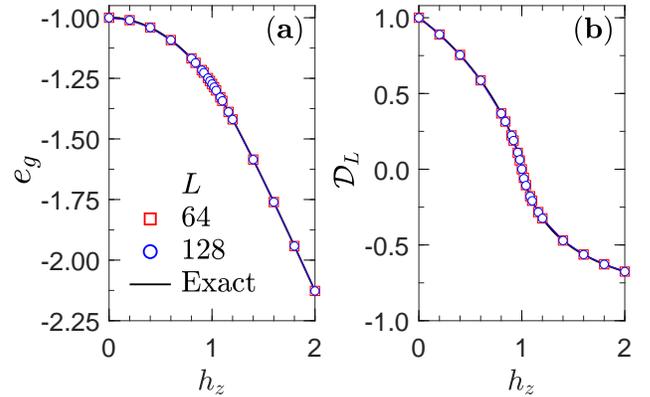}\\
\caption{(a) The ground-state energy $e_g$ of TFIM for $L$ = 64~(red square), 128~(blue circle), and TDL (black line).
(b) The same setup as (a) for DBS $\mathcal{D}_L$.}\label{FIG-TFIMIsing}
\end{figure}

For the TFIM, there is an Ising transition at $h_z = 1$ that separates the FM phase from the paramagnetic phase.
The energy per site is\cite{SMPfeuty_1970}
\begin{equation}\label{TFIM-Eg}
e_g = -\frac2\pi (1+h_z) E\left(\frac{2\sqrt{h_z}}{1+h_z}\right)
\end{equation}
where $E(\cdot)$ is the complete elliptic integral of the \textit{second} kind.
The transverse magnetization $M_z = \langle \sigma_i^z\rangle$ could also be calculated as\cite{SMMacWoj2016}
\begin{align}\label{TFIM-Mag}
    \frac{M_z}{2} =
    \left\{
    \begin{array}{ll}
        \frac{(1-h_z^2)}{\pi h_z}\left[\Pi(h_z^2; h_z) - K(h_z)\right], & h_z < 1\\
        \frac{(h_z^2-1)}{\pi h_z^2}\Pi\big(\frac{1}{h_z^2}; \frac{1}{h_z}\big), & h_z > 1
    \end{array}
    \right..
\end{align}
The correlation function $\langle \sigma_i^x\sigma_{i+1}^x\rangle = - (e_g+h_zM_z)$.
The DBS is defined as
$\mathcal{D}_L$ = $\langle \sigma_i^x\sigma_{i+1}^x\rangle - \langle \sigma_i^z\rangle$,
and the final result is presented in Fig.~\ref{FIG-TFIMIsing}.
It could be found that the DBS is smooth when crossing the critical point $h_z = 1$.
Since the Ising transition is the continuous QPT of the second order and
the KT transition is the continuous QPT of the infinite order,
we thus could arrive at the conclusion that the DBS is always smooth~(or continuous) for the continuous QPT.
As a result, the jump of DBS is a characteritic feature of the first-order QPT.

\vspace{-0.20cm}
\subsection{Commensurate-incommensurate transition}
\begin{figure}[!ht]
\centering
\vspace{-0.20cm}
\includegraphics[width=0.95\columnwidth, clip]{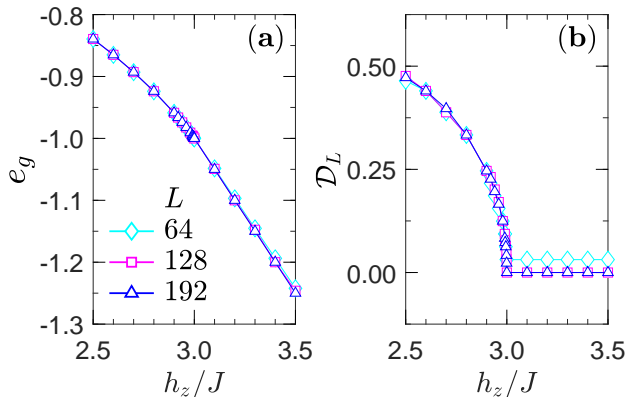}\\
\caption{(a) The ground-state energy $e_g$ for
$L$ = 64~(cyan rhombus), $L$ = 128~(magenta square), and $L$ = 192~(blue triangle).
(b) The same setup as (a) for DBS $\mathcal{D}_L$.}\label{FIG-XXZhz}
\end{figure}

We now consider the spin-$1/2$ XXZ chain under the longitudinal field.
The Hamiltonian reads as\cite{SMYangYang1966}
\begin{equation*} 
\mathcal{H} = J\sum_{i}(S_i^xS_{i+1}^x + S_i^yS_{i+1}^y + \Delta S_i^zS_{i+1}^z) - h_z\sum_{i} S_i^z
\end{equation*}
where $J = 1$ and $h_z~(\geq 0)$ is the longitudinal field.
There are two gapped phases: A ferromagnetic one at sufficiently strong fields and
an antiferromagnetic phase for $\Delta > 1$ at small fields in the full phase diagram.
Also, a massless Luttinger phase is sandwiched between the two\cite{SMSchollwockBook_2004}.
The longitudinal correlation function of the Luttinger phase
is incommensurate with sinusoidally modulated behavior\cite{SMGrenierSCetal2015}.
The transition between the ferromagnetic commensurate phase and the massless incommensurate phase,
which occurs on the line $h_{u,c}/J = 1 + \Delta$,
is an example of the Dzhaparidze-Nersesyan-Pokrovsky-Talapov universality class\cite{SMDzhNer1978,SMPokTal1979}.
We now focus on the line of $\Delta = 2$, and we define the DBS as
$\mathcal{D}_L$ = $\langle (S_i^xS_{i+1}^x + S_i^yS_{i+1}^y + 2 S_i^zS_{i+1}^z) - S_i^z \rangle_{i=L/2}$.
The final result is presented in Fig.~\ref{FIG-XXZhz}.
It could be found that the DBS is continuous when crossing the critical point $h_{z,c} = 3$.

\section{QPTs of the cross-coupled spin ladder}
\subsection{Continuous QPT at $J_{\times}$ = 0.2}
\begin{figure}[!ht]
\centering
\includegraphics[width=0.95\columnwidth, clip]{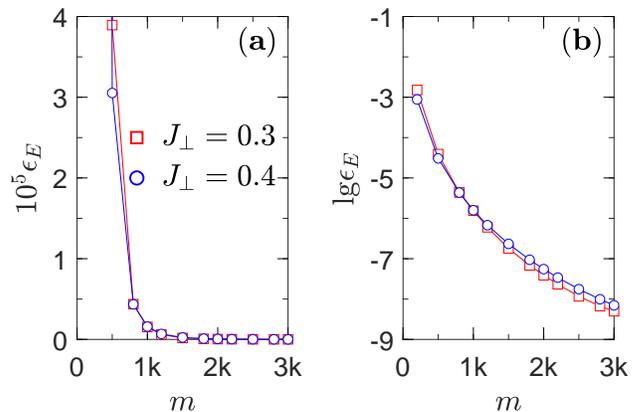}\\
\caption{Evolution of (a) the absolute error of energy $\epsilon_E$ and (b) its logarithmic form versus states kept $m$.
Here, total length of the ladder is $L = 128$ and $J_{\times} = 0.2$. 
Two different points at $J_{\perp}$ = 0.30~(red square) and 0.40~(blue circle) are selected
as the representative points in the Haldane phase and the RS phase, respectively.}\label{FIG-DMRGConverged}
\end{figure}

The fact that a direct continuous QPT occurs at $J_{\times}$ = 0.2 has been checked by several different methods.
We here revisit the problem by studying the energy gap and central charge.
Before carrying large-scale numerical calculations,
we firstly present some details about our DMRG simulations to show the convergence of our results.
In Fig.~\ref{FIG-DMRGConverged} we show the behavior of ground-state energy $E_g$ versus states kept $m$.
It could be found that as long as 2000 states are kept in the simulations
we shall reduce the absolute error of energy $\epsilon_E = E_g^{(m)} - E_g^{(\infty)}$ to seven place of decimals
where $E_g^{(m)}$ and $E_g^{(\infty)}$ represent the energy at current states kept $m$ and infinite states kept $m\to\infty$.
The truncated error of information loss here is of the order $10^{-12}$, which is fairly small.
Therefore, we keep typically 2000 states in our calculations and
4-8 sweeps are performed to ensure our results are well converged.

\begin{figure}[!ht]
\centering
\includegraphics[width=0.95\columnwidth, clip]{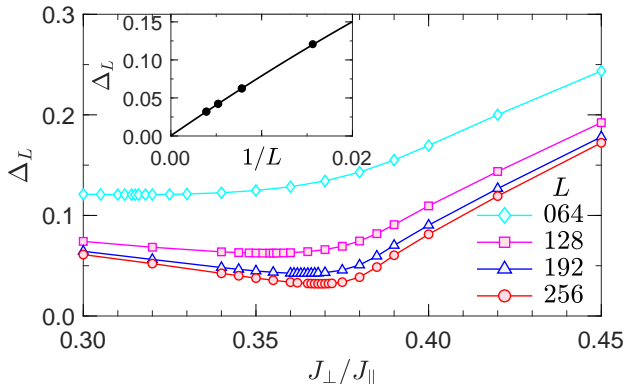}\\
\caption{The Haldane gap $\Delta_L$ at fixed $J_{\perp} = 0.2$ for different chain length $L$.
The inset shows a linear extrapolation of $\Delta_L$ to TDL.}\label{FIG-JX020GAP}
\end{figure}

We now turn to calculate the energy gap $\Delta$ at the critical point and the central charge $c$, if any.
In the open boundary condition, the Haldane phase has a four-fold degenerate ground state due to the edge modes.
The Haldane gap $\Delta_L$ is thus defined as the difference of ground-state energy
in the $S^z_{\textrm{tot}} = 2$ and $S^z_{\textrm{tot}} = 0$ subspaces, i.e.,
$\Delta_L = E_g\big(S^z_{\textrm{tot}} = 2\big) - E_g\big(S^z_{\textrm{tot}} = 0\big)$.
The DMRG result of the Haldane gap $\Delta_L$ is shown in Fig.~\ref{FIG-JX020GAP} and
it could be seen that local minima exist near the critical point in the finite systems.
Such minimal gaps obey a linear scaling formula and the gap in the TDL is 0.0001(2),
which could be regarded as zero within the numerical precision.
The vanishing gap at the critical point is a typical signal for a continuous QPT,
and the linear fitting suggests that the critical point could be described by the conformal field theory.

\begin{figure}[!ht]
\centering
\includegraphics[width=0.95\columnwidth, clip]{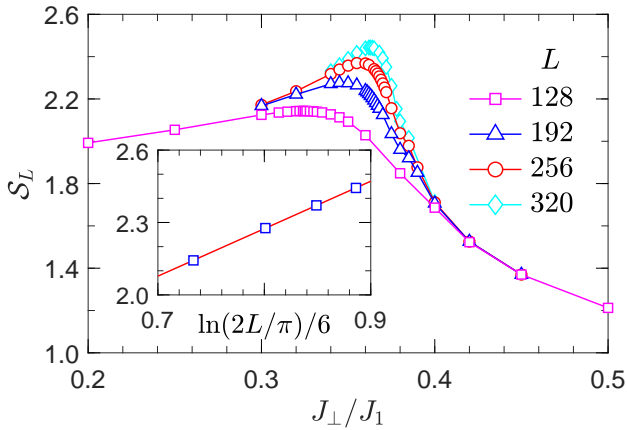}\\
\caption{The von Neumann entropy $\mathcal{S}_L$ at fixed $J_{\perp} = 0.2$ for different chain length $L$.
The inset shows a proper fitting of the maxima of $\mathcal{S}_L$ to extract the central charge $c$ and
the best estimated value is $c = 1.98(4)$.}\label{FIG-JX020vNE}
\end{figure}

Meanwhile, we also calculate the von Neumann entropy $\mathcal{S}_L$ and the result is presented in Fig.~\ref{FIG-JX020vNE}.
A bump could be spotted near the critical point and is consistent with a continuous QPT.
It is well established that for the critical system under the OBC the vNE obeys the following formula,
\begin{equation}\label{SM-vNECC}
\mathcal{S}_L(l) = \frac{c}{6}\ln\Big(\frac{2L}{\pi}\sin\Big(\frac{\pi l}{L}\Big)\Big) + c'
\end{equation}
where $c$ is the central charge.
The result in Fig.~\ref{FIG-JX020vNE} is a special case where $l = L/2$.
It could be seen from the inset that the central charge $c \simeq 2$.
It's worth mentioning that the central charge at the decoupling limit where $J_{\perp} = 0$ is also equal to 2.

\subsection{First-Order QPT at $J_{\times}$ = 0.4}
\begin{figure}[!ht]
\centering
\includegraphics[width=0.95\columnwidth, clip]{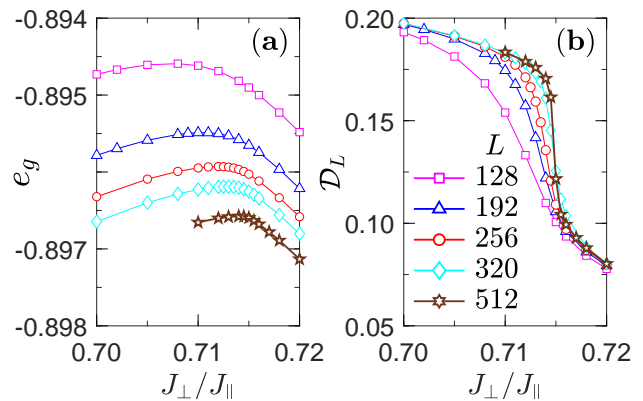}\\
\caption{(a) The ground-state energy $e_g$ of the cross-coupled spin ladder for
$L$ = 128~(magenta square), $L$ = 192~(blue triangle), $L$ = 256~(red circle), $L$ = 320~(cyan rhombus), and $L$ = 512~(auburn star)
at fixed $J_{\times} = 0.4$.
(b) The same setup as (a) for DBS $\mathcal{D}_L$.}\label{FIG-JX040}
\end{figure}

Whereas the DBS of the cross-coupled spin ladder seems to be smooth when $J_{\times}$ = 0.4~(see Fig.~3 in the main text),
we want to convince the readers that there is a jump actually.
For this purpose we calculate the DBS of a longer ladder whose length $L$ is up to 512.
As shown in Fig~\ref{FIG-JX040}, we find that the energy curves are more and more screwy and a kink is expected for an infinite system.
Likewise, for the DBS, it is very smooth for small sizes and a jump could be spotted when $L = 512$.
Therefore, we think that the transition at $J_{\times}$ = 0.4 is still of first order and there is a jump of DBS correspondingly.

\begin{figure}[!ht]
\centering
\includegraphics[width=0.95\columnwidth, clip]{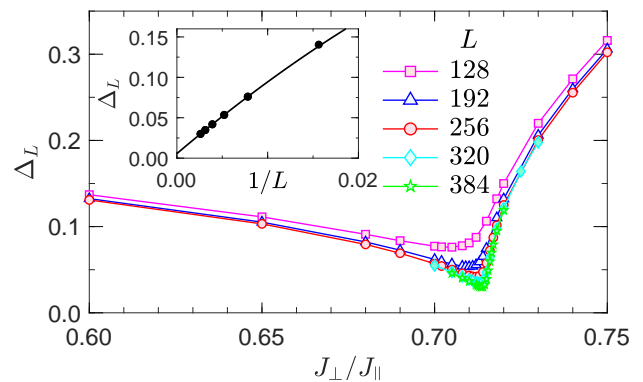}\\
\caption{The Haldane gap $\Delta_L$ at fixed $J_{\times} = 0.4$ for different chain length $L$.
The inset shows a quadratic extrapolation of $\Delta_L$ to TDL.}\label{FIG-JX040GAP}
\end{figure}

Likewise, we calculate the Haldane gap as before and find that it is finite of ~0.0056(8) in the TDL.
This further confirms that the transition at $J_{\times}$ = 0.4 is of first order.

\subsection{Transition points and the Haldane gaps}
The transition points are obtained by the joint analysis of
the maxima of vNE $\mathcal{S}_L$ and minima of Haldane gap $\Delta_{L}$ for the continuous QPT,
and by the jump of DBS for the first-order one.
The final results are presented in Tab.~\ref{SMTab-CCSLGAP}.
\vspace{-0.10cm}
\begin{table}[!ht]
\caption{\label{SMTab-CCSLGAP}
Transition point $J_{\perp, t}$ and triplet gap $\Delta_{\textrm{m}}$ of Haldane phase thereof for several selected $J_{\times}$.}
\begin{ruledtabular}
\begin{tabular}{c c c | c c c}
$J_{\times}$ & $J_{\perp, t}$ & $\Delta_{\textrm{m}}$ & $J_{\times}$ & $J_{\perp, t}$ & $\Delta_{\textrm{m}}$ \\
\colrule
0.20         & 0.3826(3)      & 0.0001(2)             & 0.45         & 0.7918(6)       & 0.0158(2)              \\
0.30         & 0.5576(5)      & 0.0004(5)             & 0.50         & 0.8625(5)       & 0.0355(3)              \\
0.40         & 0.7168(3)      & 0.0037(9)             & 0.60         & 0.9990(5)       & 0.1265(5)              \\
\end{tabular}
\end{ruledtabular}
\end{table}

\section{Critical exponents of the spin-1/2 chain}
\begin{figure}[!ht]
\centering
\includegraphics[width=0.95\columnwidth, clip]{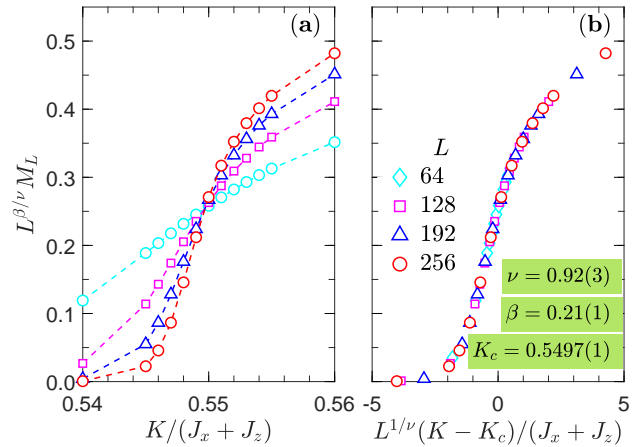}\\
\caption{FSS analysis of the VBS order parameter $M_L$ for $\delta = 0.5$ as a function of $K/(J_x+J_z)$.
The best fitting suggests $K_c$ = 0.5497(1) and critical exponents $\beta$ = 0.21(1) and $\nu$ = 0.92(3).}\label{FIG-FSSDlt050}
\end{figure}

In this section we pick up another set of parameters shown in Eq.~(3) of the main text.
Here we fix $\tilde{J} = (J_z+J_x)/2 = 1$ as the energy unit
and introduce an anisotropic parameter $\delta = (J_z-J_x)/(J_z+J_x)$.
Let $\delta = 0.5$ and the finite-size scaling~(FSS) analysis of
the corresponding VBS order parameter $M_L$ are shown in Fig.~\ref{FIG-FSSDlt050}.
After a careful analysis, we find that the critical point $K_c$ = 0.5497(1)
and the associated critical exponents $\beta$ = 0.21(1) and $\nu$ = 0.92(3).
Those results are in fairly consistent with other group\cite{SMRobertsJM2019}.
Together with the result shown in the main text,
we can conclude that the critical exponents vary along the locus of the phase boundary.


\begin{thebibliography}{99}
\bibitem{SachdevBook_2011}
  S. Sachdev, {\em\textit{Quantum Phase Transitions}}
  \newblock (Cambridge University Press, Cambridge, 2011).

\bibitem{LMMBook_2011}
  C. Lacroix, P. Mendels, and F. Mila, {\em\textit{Introduction to Frustrated Magnetism}}
  \newblock (Springer-Verlag Berlin Heidelberg, 2011).

\bibitem{WenBook_2019}
  B. Zeng, X. Chen, D.-L. Zhou, and X.-G. Wen, {\em\textit{Quantum Information Meets Quantum Matter}}
  \newblock (Springer-Verlag New York, 2019).



\bibitem{KT_1973}
  J.~M. Kosterlitz, and D.~J. Thouless,
  J. Phys. C \textbf{6}, 1181 (1973).

\bibitem{TsuiSG_1982}
  D. C. Tsui, H. L. Stormer, and A. C. Gossard,
  Phys. Rev. Lett. \textbf{48}, 1559 (1982).

\bibitem{Haldane_1983}
  F.~D.~M. Haldane,
  Phys.~Rev.~Lett. \textbf{50}, 1153 (1983).

\bibitem{SenthilVBetal2004}
  T. Senthil, A. Vishwanath, L. Balents, S. Sachdev, and M. P. A. Fisher,
  Science \textbf{303}, 1490 (2004).

\bibitem{SenthilBSetal2004}
  T. Senthil, L. Balents, S. Sachdev, A. Vishwanath, and M. P. A. Fisher,
  Phys. Rev. B \textbf{70}, 144407 (2004).

\bibitem{Sandvik2007}
  A.~W. Sandvik,
  Phys. Rev. Lett. \textbf{98}, 227202 (2007).

\bibitem{ShaoGS2016}
  H. Shao, W. Guo, A.~W. Sandvik,
  Science \textbf{352}, 213 (2016).


\bibitem{PollmannTBetal_2010}
  F.~Pollmann, A.~M. Turner, E.~Berg, and M.~Oshikawa,
  Phys.~Rev.~B {\bf 81}, 064439 (2010).

\bibitem{FaureBCVO_2018}
  Q. Faure, S. Takayoshi, S. Petit, V. Simonet, S. Raymond, L.-P. Regnault, M. Boehm, J.~S. White, M. M{\aa}nsson, C. R{\"u}egg, P. Lejay, B. Canals, T. Lorenz, S.~C. Furuya, T. Giamarchi, and B. Grenier,
  Nat. Phys. \textbf{14}, 716 (2018).

\bibitem{Wang_2000}
  X.~Wang,
  Modern Phys.~Lett.~B {\bf 14}, 327 (2000).

\bibitem{AmaricciBCetal_2015}
  A. Amaricci, J. C. Budich, M. Capone, B. Trauzettel, and G. Sangiovanni,
  Phys.~Rev.~Lett. \textbf{114}, 185701 (2015).

\bibitem{BarbarinoSB_2019}
  S. Barbarino, G. Sangiovanni, and J.~C. Budich,
  Phys.~Rev.~B {\bf 99}, 075158 (2019).

\bibitem{RoyGS_2016}
  B. Roy, P. Goswami, and J.~D. Sau,
  Phys.~Rev.~B {\bf 94}, 041101(R) (2016).

\bibitem{Sandvik2010}
  A.~W. Sandvik,
  Phys. Rev. Lett. \textbf{104}, 177201 (2010).

\bibitem{ChenHDetal2013}
  K. Chen, Y. Huang, Y. Deng, A. B. Kuklov, N. V. Prokof'ev, and B. V. Svistunov,
  Phys. Rev. Lett. \textbf{110}, 185701 (2013).

\bibitem{IaizziDS2018}
  A. Iaizzi, K. Damle, and A. W. Sandvik,
  Phys. Rev. B \textbf{98}, 064405 (2018).

\bibitem{FinkDVetal2017}
  J. M. Fink, A. Dombi, A. Vukics, A. Wallraff, and P. Domokos,
  Phys.~Rev.~X {\bf 7}, 011012 (2017).

\bibitem{LaumannMSetal_2012}
  C. R. Laumann, R. Moessner, A. Scardicchio, and S. L. Sondhi,
  Phys.~Rev.~Lett. {\bf 109}, 030502 (2012).

\bibitem{MuellerJJ_2014}
  M. Mueller, W. Janke, and D.~A. Johnston,
  Phys.~Rev.~Lett. {\bf 112}, 200601 (2014).

\bibitem{CampostriniNPetal_2014}
  M. Campostrini, J. Nespolo, A. Pelissetto, and E. Vicari,
  Phys.~Rev.~Lett. {\bf 113}, 070402 (2014).

\bibitem{YusteCCetal_2018}
  A. Yuste, C. Cartwright, G. De Chiara, and A. Sanpera,
  New J. Phys. {\bf 20}, 043006 (2018).

\bibitem{RossiniVicari_2018}
  D. Rossini and E. Vicari,
  Phys.~Rev.~E {\bf 98}, 062137 (2018).

\bibitem{SuppMat}
  For details see the Supplemental Material at [url],
  which includes the DBS of continuous QPTs, QPTs of the cross-coupled spin ladder,
  and the critical exponents of the spin-$1/2$ chain with DQCP.

\bibitem{White_1992}
  S.~R. White,
  Phys.~Rev.~Lett.~{\bf 69}, 2863 (1992).

\bibitem{White_1993}
  S.~R. White,
  Phys.~Rev.~B {\bf 48}, 10345 (1993).

\bibitem{PeschelBook_1999}
  I.~Peschel, X.~Q. Wang, M.~Kaulke, and K.~Hallberg,
  {\em Density-matrix renormalization}.
  \newblock (Springer Berlin Heidelberg, 1999).

\bibitem{Schollwoeck_2005}
  U.~Schollw{\"o}ck,
  Rev.~Mod.~Phys.~{\bf 77}, 259 (2005).


\bibitem{SchollwockBook_2004}
  U. Schollw{\"o}ck, J. Richter, D. J. J. Farnell, and R. F. Bishop,
  {\em Quantum Magnetism}.
  \newblock (Springer Berlin Heidelberg, 2004).

\bibitem{YangYang1966}
  C.~N. Yang and C.~P. Yang,
  Phys.~Rev. \textbf{150}, 321 (1966).

\bibitem{ShiroishiTakahashi2015}
  M. Shiroishi and M. Takahashi,
  J. Phys. Soc. Jpn. \textbf{74}, 47 (2015).

\bibitem{CloizeauxGaudin1966}
  J. D. Cloizeaux and M. Gaudin,
  J. Math. Phys. \textbf{7}, 1384 (1966).

\bibitem{BortzSS2007}
  M. Bortz, J. Sato, and M. Shiroishi,
  J. Phys. A: Math. Theor. \textbf{40}, 4253 (2007).

\bibitem{JimboMiwa1996}
  M. Jimbo and T. Miwa,
  J. Phys. A \textbf{29}, 2923 (1996).

\bibitem{KatoSTetal2003}
  G. Kato, M. Shiroishi, M. Takahashi, and K. Sakai,
  J. Phys. A \textbf{36}, L337 (2003).

\bibitem{BanchiCV2009}
  L. Banchi, F. Colomo, and P. Verrucchi,
  Phys.~Rev.~A {\bf 80}, 022341 (2009).

\bibitem{ErcolessiEFetal2011}
  E. Ercolessi, S. Evangelisti, F. Franchini, and F. Ravanin,
  Phys.~Rev.~B {\bf 83}, 012402 (2011).


\bibitem{AlbaHL2012}
  V. Alba, M. Haque, and A.~M L{\"a}uchli,
  J. Stat. Mech. {\bf 2012}, P08011 (2012).

\bibitem{StasinskaRPetal2014}
  J. Stasinska, B. Rogers, M. Paternostro, G. De Chiara, and A. Sanpera,
  Phys.~Rev.~A {\bf 89}, 032330 (2014).

\bibitem{Pfeuty_1970}
  P. Pfeuty,
  Ann. Phys. (N.Y.)~{\bf 57}, 79 (1970).

\bibitem{DuttaACetal_2015}
  A.~Dutta, G.~Aeppli, B.~K. Chakrabarti, U. Divakaran, T.~F. Rosenbaum, and D. Sen,
  {\em Quantum Phase Transitions in Transverse Field Spin Models}.
  \newblock (Cambridge University Press, 2015).

\bibitem{ColdeaTWetal_2010}
  R. Coldea, D.~A. Tennant, E.~M. Wheeler, E. Wawrzynska, D. Prabhakaran, M. Telling, K. Habicht, P. Smeibidl, and K. Kiefer,
  Science~{\bf 327}, 177 (2010).

\bibitem{Zamolodchikov_1989}
  A.~B. Zamolodchikov,
  Int. J. Mod. Phys. A~{\bf 4}, 4235 (1989).

\bibitem{ODK_2003}
  A. A. Ovchinnikov, D. V. Dmitriev, and V. Ya. Krivnov, and V. O. Cheranovskii,
  Phys.~Rev.~B {\bf 68}, 214406 (2003).

\bibitem{AtasBogomolny_2017}
  Y.~Y. Atas and E. Bogomolny,
  J. Phys. A: Math. Theor. {\bf 50}, 385102 (2017).


\bibitem{BednorzMuller_1986}
  J. G. Bednorz and K. A. M{\"u}ller,
  Z. Phys. B {\bf 64}, 189 (1986).

\bibitem{Xian_1995}
  Y.~Xian,
  Phys.~Rev.~B {\bf 52}, 12485 (1995).

\bibitem{WesselNMH_2017}
  S. Wessel, B. Normand, F. Mila, and A. Honecker,
  SciPost Phys. {\bf 3}, 005 (2017).

\bibitem{Dagotto_1996}
  E.~Dagotto and T.~M. Rice,
  Science {\bf 271}, 618 (1996).

\bibitem{WhiteLadder_1996}
  S. R. White,
  Phys.~Rev.~B {\bf 53}, 52 (1996).

\bibitem{Vekua_2006}
  T.~Vekua and A.~Honecker,
  Phys.~Rev.~B {\bf 73}, 214427 (2006).

\bibitem{Metavitsiadis_2017}
  A.~Metavitsiadis and S.~Eggert,
  Phys.~Rev.~B {\bf 95}, 144415 (2017).

\bibitem{HijiiKN_2005}
  K. Hijii, A. Kitazawa, and K. Nomura,
  Phys.~Rev.~B {\bf 72}, 014449 (2005).

\bibitem{WeihongKO_1998}
  Z. Weihong, V. Kotov, and J. Oitmaa,
  Phys.~Rev.~B {\bf 57}, 11439 (1998).

\bibitem{KimFSetal_2000}
  E. H. Kim, G. F{\'a}th, J. S{\'o}lyom, and D. J. Scalapino,
  Phys.~Rev.~B {\bf 62}, 14965 (2000).

\bibitem{KimLS_2008}
  E. H. Kim, {\"O}. Legeza, and J. S{\'o}lyom,
  Phys.~Rev.~B {\bf 77}, 205121 (2008).


\bibitem{Starykh_2004}
  O.~A. Starykh and L.~Balents,
  Phys.~Rev.~Lett.~{\bf 93}, 127202 (2004).

\bibitem{Liu_2008}
  G.-H. Liu, H.-L. Wang, and G.-S. Tian,
  Phys.~Rev.~B {\bf 77}, 214418 (2008).

\bibitem{Li_2012}
  Y.-C. Li and H.-Q. Lin,
  New J.~Phys.~{\bf 14}, 063019 (2012).

\bibitem{Hung_2006}
  H.-H. Hung, C.-D. Gong, Y.-C. Chen, and M.-F. Yang,
  Phys.~Rev.~B {\bf 73}, 224433 (2006).

\bibitem{Kim_2008}
  E.~H. Kim, {\"O}.~Legeza, and J.~S{\'o}lyom,
  Phys.~Rev.~B {\bf 77}, 205121 (2008).

\bibitem{Hikihara_2010}
  T.~Hikihara and O.~A. Starykh,
  Phys.~Rev.~B {\bf 81}, 064432 (2010).

\bibitem{Barcza_2012}
  G.~Barcza, {\"O}.~Legeza, R.~M. Noack, and J.~S{\'o}lyom,
  Phys.~Rev.~B {\bf 86}, 075133 (2012).

\bibitem{ChenCZetal_2016}
  X.-H. Chen, S. Y. Cho, H.-Q. Zhou, and M. T. Batchelor,
  J. Korean Phys. Soc. \textbf{68}, 1114 (2016).

\bibitem{JiangMotrunich2019}
  S. Jiang and O. Motrunich,
  Phys. Rev. B \textbf{99}, 075103 (2019).


\bibitem{RobertsJiangMotrunich2019}
  B. Roberts, S. Jiang, and O. I. Motrunich,
  Phys. Rev. B \textbf{99}, 165143 (2019).

\bibitem{HuangLYetal2019}
  R.-Z. Huang, D.-C. Lu, Y.-Z. You, Z.~Y. Meng, and T. Xiang,
  arXiv:1904.00021 (2019).

\bibitem{MudryFMetal2019}
  C. Mudry, A. Furusaki, T. Morimoto, and T. Hikihara,
  Phys. Rev. B \textbf{99}, 205153 (2019).

\bibitem{MajumdarGhosh_1969}
  C.~K. Majumdar and D.~K. Ghosh, J.~Math.~Phys.~{\bf 10}, 1388 (1969); \textit{ibid}, {\bf 10}, 1399 (1969).

\bibitem{FrancescoMSBook_1997}
  P.~Di~Francesco, P.~Mathieu, and D.~Senechal, {\em \textit{Conformal field theory}}
  \newblock (Springer, New York, 1997).

%

\bibitem{CalabreseCardy2004}
  P. Calabrese, and J. Cardy,
  J. Stat. Mech.: Theor. Exp. \textbf{2004}, P06002 (2004).

\bibitem{BarberBook_1983}
  M. N. Barber, {\em \textit{Phase Transitions and Critical Phenomena}} Vol. 8 (eds C. Domb and J. L. Leibovitz)
  \newblock (Academic, London, 1983).

\bibitem{SunWeiKou2019}
  G. Sun, B.-B. Wei, and S.-P. Kou,
  arXiv:1906.03850 (2019).


\end{thebibliography}

\begin{thebibliography}{99}%
\makeatletter

\bibitem{SMKatoSTetal2003}
  G. Kato, M. Shiroishi, M. Takahashi, and K. Sakai,
  J. Phys. A \textbf{36}, L337 (2003).

\bibitem{SMTakahashiKS2003}
  M. Takahashi, G. Kato, and M. Shiroishi,
  J. Phys. Soc. Jpn. \textbf{73}, 245 (2004).

\bibitem{SMPfeuty_1970}
  P. Pfeuty,
  Ann. Phys. (N.Y.)~{\bf 57}, 79 (1970).

\bibitem{SMMacWoj2016}
  T. Maciazek and J. Wojtkiewicz,
  Physica A~{\bf 441}, 131 (2016).

\bibitem{SMYangYang1966}
  C.~N. Yang and C.~P. Yang,
  Phys.~Rev. \textbf{150}, 321 (1966).

\bibitem{SMSchollwockBook_2004}
  U. Schollw{\"o}ck, J. Richter, D. J. J. Farnell, and R. F. Bishop,
  {\em Quantum Magnetism}.
  \newblock (Springer Berlin Heidelberg, 2004).

\bibitem{SMGrenierSCetal2015}
  B. Grenier, V. Simonet, B. Canals, and P. Lejay, M. Klanjsek, M. Horvatic, and C. Berthier,
  Phys. Rev. B \textbf{92}, 134416 (2015).

\bibitem{SMDzhNer1978}
  G.~I. Dzhaparidze, A.~A. Nersesyan,
  JETP Lett. \textbf{27}, 334 (1978).

\bibitem{SMPokTal1979}
  V.~L. Pokrovsky, A.~L. Talapov,
  Phys. Rev. Lett. \textbf{42}, 65 (1979).

\bibitem{SMRobertsJM2019}
  B. Roberts, S. Jiang, and O. I. Motrunich,
  Phys. Rev. B \textbf{99}, 165143 (2019).

\end{thebibliography}
%


\end{document}